# Argon Cluster-Ion Sputter Yield: Molecular Dynamics Simulations on Silicon and an Equation for Estimating Total Sputter Yield

Peter J Cumpson[1]*, Mieszko Jaskiewicz[1] and Woo Kyun Kim[2]

(1) NEXUS laboratory, School of Engineering, Newcastle University, Newcastle upon Tyne, NE1 7RU, UK

(2) Department of Mechanical and Materials Engineering, University of Cincinnati, USA

* Author to whom correspondence should be addressed. Email: peter.cumpson@ncl.ac.uk

ABSTRACT

Argon Gas Cluster-Ion Beam sources have become widely-used on x-ray photoelectron spectroscopy (XPS) and secondary ion mass spectrometry (SIMS) instruments in recent years, but there is little reference data on sputter yields in the literature as yet. Total sputter yield reference data is needed in order to plan, and later calibrate the depth scale, of XPS or SIMS depth profiles. We previously published a semi-empirical "Threshold" equation for estimating cluster total sputter yield from the energy-per-atom of the cluster and the effective monatomic sputter threshold of the material. This has already been shown to agree extremely well with sputter yield measurements on a range of organic and inorganic materials for clusters of around a thousand atoms. Here we use the molecular dynamics (MD) approach to explore a wider range of energy and cluster size than is easy to do experimentally to high precision. We have performed MD simulations using the *"Large-scale Atomic/Molecular Massively Parallel Simulator"* (LAMMPS) parallel MD code on high-performance computer (HPC) systems at Cincinnati and Newcastle. We performed 1,150 simulations of individual collisions with a silicon (100) surface as an archetypal inorganic substrate, for cluster sizes between 30 and 3,000 argon atoms and energies in the range 5 to 40eV per atom. This corresponds to the most important regime for experimental cluster depth-profiling in SIMS and XPS. Our MD results show a dependence on cluster size as well as energy-per-atom. Using the exponent previously suggested by Paruch *et al*, we modified the Threshold model equation published previously to take this into account. The modified Threshold equation fits all our MD results extremely well, building on its success in fitting experimental sputter yield measurements.

1. **Introduction**

Ion beam sputtering has been used for many years in conjunction with surface analytical techniques to provide profiles of composition as a function of depth. Surface layers are removed in a succession of sputtering steps, and x-ray photoelectron spectroscopy (XPS) or secondary ion mass spectrometry (SIMS) are used to analyse the surface revealed. Recently gas cluster ion beam (GCIB) columns have begun to be commercially available on surface analysis instruments and are now quite widely-used. Argon cluster-ion sources were originally developed for semiconductor processing[1],[2] and advanced coatings[3], and subsequently the use of these sources for analytical applications was pioneered in SIMS[4,5]. They are equally useful in XPS[6],[7], where for example sputter depth profiling of organic materials[8], such as semiconductors[9] can give access to defect and band bending measurements at interfaces. Heritage materials can be analysed after removal of organic films[10], and many applications can be found in the field of inorganic analysis[11],[12]. There is still an absence of sputter yield reference data, made worse by the fact that unlike monatomic sputter profiling, cluster ions used in XPS or SIMS are typically in a regime where the energy of each atom in the cluster is close to the monatomic sputter threshold[13], making sputter yield rather more complex. Above the threshold one should expect a high yield, whereas below it, sputtering can still occur *via* the collective effects of many cluster atoms imparting an energy above the threshold. Therefore the distribution of energy amongst these cluster atoms at the point of impact is important to the estimation of total sputter yield, even though no sharp threshold feature is seen. This is a "many body" problem, and therefore it is no surprise that (in the absence of good reference data) Molecular Dynamics (MD) simulations have been extremely useful over the last decade in elucidating the details of cluster-ion sputtering processes[14,15,16].

A few years ago we proposed the following "Threshold Model"[17] semi-empirical equation for estimating total sputter yield $Y$ as a function of the average kinetic energy-per-atom, $\varepsilon$, of the incident argon clusters;

$$Y(\epsilon) = A\, n\epsilon \left[1 + \mathrm{erf}\left(\frac{\epsilon - U}{s}\right)\right]$$

(1)

This is based on a very simple model of the distribution of energy on impact, where $U$ represents the atomic sputtering threshold in eV, $s$ the range of energy of atoms within the cluster immediately after impact, again in eV, and $n$ is the number of argon atoms in the incident cluster. $A$ is a constant for a given sputtered material. Even though there is undoubtedly a very complex range of effects taking place during GCIB sputtering, Eqn (1) was found to be an excellent fit[17] to experimental measurements of total sputter yield for $n \approx 1000$. To understand the parameter dependence of the yield more completely, and in particular to investigate the effect of cluster size $n$, molecular dynamics simulation is an extremely useful tool since it allows to directly observe the atomic-level interactions during impact. Therefore in the present work we have performed a large number of MD simulations using the *"Large-scale Atomic/Molecular Massively Parallel Simulator (LAMMPS)"*[18] parallel MD code. These simulations were performed on high-performance computer (HPC) clusters at Cincinnati and Newcastle. LAMMPS can model systems that include millions of atoms with ease. It is flexible and allows for definition of interactions between many different types of particles, making it ideal for sputtering simulations. Next we consider in detail the model created in LAMMPS, firstly the simulation domain and then the important issue of choosing valid interaction potentials between the atom types taking part in any sputtering process.

## 2. Simulation Method

**2.1 Domain**

The size of the simulation domain and the boundary conditions are critical. Ideally, the domain size would be large enough for argon cluster impact to have no effect on atoms close to the target boundary. However, this is impractical due to the long simulation time that would be needed, but we can achieve very nearly this ideal condition by choosing to simulate a large cell and using periodic boundary conditions. The size of the simulation has been based on previous sputtering MD simulations, which have been chosen to minimise the ejection of particles due to edge effects impacting on the dynamical events[19,20]. In our simulations the Si (100) substrate is a cube and contains 407,962 silicon atoms; it has dimensions of 200Å×200Å×200Å. This is a single substance; we make no attempt to model the native oxide one would expect on a real Si surface exposed to air. This therefore reflects the situation in ultra-high vacuum (UHV) analysis of a clean Si surface. The simulation box is illustrated in *Figure 1.* All the figures illustrating Si and Ar atoms in this paper were prepared using Open Visualization Tool (OVITO)[21].

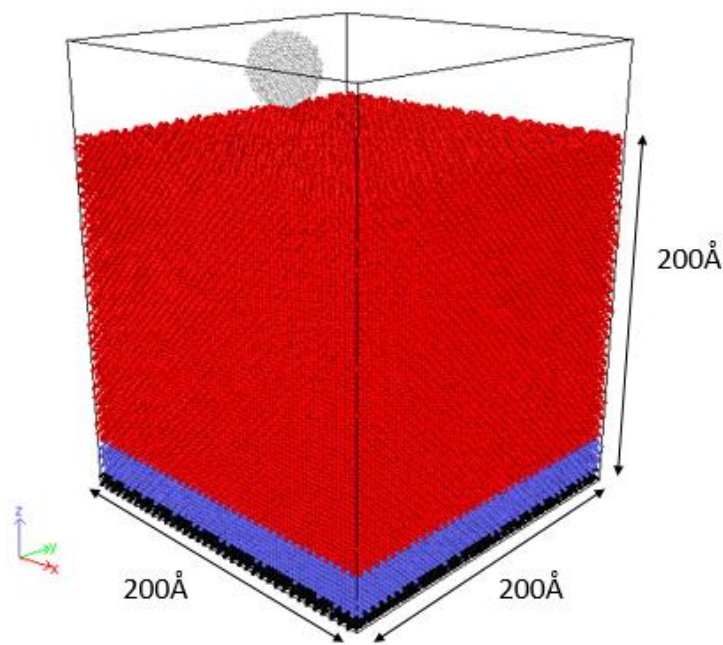

Figure 1 Simulation volume used in our LAMMPS simulation. Note the argon cluster (in grey) about to impact the Si surface, and coloured silicon atoms to indicate those that are fixed (black), a thermal bath (blue) and fully Newtonian-mechanics simulated (red).

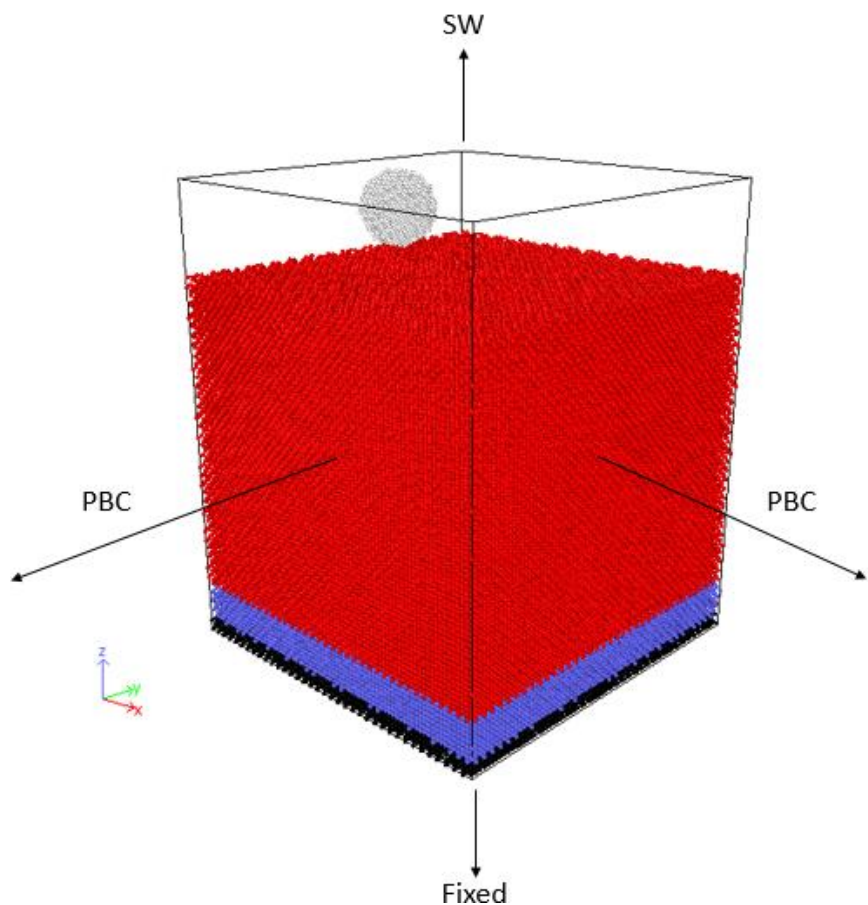

Figure 2 The boundary condition used to define global simulation box, where PBC refers to period boundary conditions and SW refers to shrink-wrapped. The fixed boundary (in black) is composed of immobile atoms

Periodic boundaries are used in the directions perpendicular to the z-axis to remove the surface effect, as shown in Figure 2.

## 2.2 Constraints

A constraint is applied to the system at each time step. It can be used to control temperature, enforce boundary conditions and apply constraint forces to atoms.

The silicon structure is separated into 3 layers, shown in Fig 1, each layer being subject to different constraints. The bottom layer (black in Fig 1) is used to prevent the rigid-body translational motion of the silicon substrate which can be imparted from the impact with the ion cluster. Therefore, all the atoms in this 0.7nm thick layer are fixed and do not move. The atoms in the top layer (red in Fig 1) are governed by Newtonian mechanics and are free to move without any constraints. To improve the accuracy of this finite-size simulation domain, a 2nm layer which acts as a thermal bath has been used[22], and is shown in blue in Fig 1. It removes the excess kinetic energy from the system, which is normally dissipated throughout the material during SIMS and XPS experiments. The thermal bath also prevents pressure waves reflecting from system boundaries[23]. The thermal bath is kept at a constant temperature $T$ of 300K. The instantaneous temperature of the system, $T_{inst}$ (defined in Eqn (2)), does actually fluctuate during the simulation, but it is important that the canonical ensemble is approximated using a proper thermostat, i.e., $< T_{inst} > = T$.

$$T_{inst} = \frac{2K(\mathbf{p})}{3Nk_B} = \frac{2}{3Nk_B}\left(\frac{\|\mathbf{p}_1\|^2}{2m_1} + \cdots \frac{\|\mathbf{p}_N\|^2}{2m_N}\right)$$

(2)

Where $K$ is the kinetic energy as a function of atom momenta $\mathbf{p}$, $m$ is the mass of each particle, $N$ is the number of atoms in the system, and $k_b$ is the Boltzmann constant. Similar to other sputtering simulations, a Langevin thermostat has been chosen to control the temperature[24],[22], where a damping term and a random force are introduced to the equation of motion to reproduce the canonical ensemble distribution of the positions and velocities of atoms in the system. The atoms in the thermal bath region are governed by the following equation:

$$m_i\ddot{\mathbf{r}}_i = -\frac{\partial V}{\partial \mathbf{r}_i} - \gamma m_i \dot{\mathbf{r}}_i + \mathbf{R}_i$$

(3)

Where $V$ is the interatomic potential, the $\gamma$ is the Langevin friction coefficient and $\mathbf{R}_i$ is the random force. The friction coefficient controls the convergence of the system.

### 2.3 Interatomic Potentials

Our simulation comprises Ar and Si atoms, and therefore three interatomic potentials must be used to describe interactions between all the different atoms in the simulation domain. Lennard-Jones potential models the interactions between Ar atoms. The high-energy collision between Ar and Si atoms is represented by potentials chosen carefully according to the properties of the atoms and the energy range in question. The detailed selection and parametrisation of these potentials is described in the Appendix.

### 2.4 Initial Conditions

In MD simulations all initial positions and velocities need to be specified. The velocity of the incident Ar atoms is determined by the kinetic energy per atom and the angle of incidence. For the simulations conducted, the kinetic energy ranges from 5eV to 40eV per atom. The angle of incidence is kept constant at 45°, commonly used for sputtering in real XPS and SIMS analytical configurations. Thus the velocity magnitude can be calculated using:

$$v = \sqrt{\frac{2\varepsilon}{m}}$$

(4)

Where $\varepsilon$ is the energy per atom, and m is the mass of an Ar atom, equal to $6.63 \times 10^{-26} kg$. To obtain a true representation of sputtering yield, the cluster must be aimed at many different positions on the surface[25],[26]. In our work this was achieved by adjusting the initial position of the cluster at random by a few nanometres in x and y. At *t*=0 there should be no interaction between the Ar and Si atoms. Consequently, the initial height of the cluster must be greater than the largest cut-off radius specified in the simulation, equal to 8Å. A safety margin of 2Å is added, meaning that the Ar atoms are at least 10Å away from the target surface. Due to lack of information on the structure of the clusters formed in SIMS simulations, a spherical cluster is assumed, with Ar atoms arranged in a face centred cubic structure oriented at 45° with respect to the global axis.

The silicon atoms are arranged in a crystal diamond cubic structure. Since total momentum is conserved over the course of the simulation, the initial momentum must also be set to zero. Velocities are assigned to the free (red in Fig 1) and thermal bath (blue in Fig 1) regions only. They must satisfy the canonical probability distribution, so a Gaussian distribution is used to generate the initial random velocities. A subsequent equilibration simulation is used to set a target temperature of 300K. The Langevin dynamics described above is applied to all but the fixed layer in the substrate. The simulation should therefore produce microstates which exactly match the canonical ensemble. The equilibration simulation is run for 100ps. We confirmed that the canonical temperature probability density distribution is satisfied and the temperature of the substrate after 100ps is very close to 300K. Once the equilibration simulation is completed, the sputtering simulation is performed with the Langevin dynamics off in the newtonian region (red in Fig. 1), i.e., the Langevin dynamics is applied only to the thermal bath region (blue in Fig. 1).

3. **Results and Discussion**

3.1 **Qualitative discussion of time evolution of individual sputtering events**

The energy of the Ar cluster in these sputtering simulations is chosen between 5eV/atom to 40eV/atom, as that is the range of practical use for surface analysis. The sputtering process for $Ar_{1000}$ with the energy of 20eV/atom is shown in Fig 3 at a series of six time steps. Sputtered Si atoms are seen to leave the surface and enter the vacuum region.

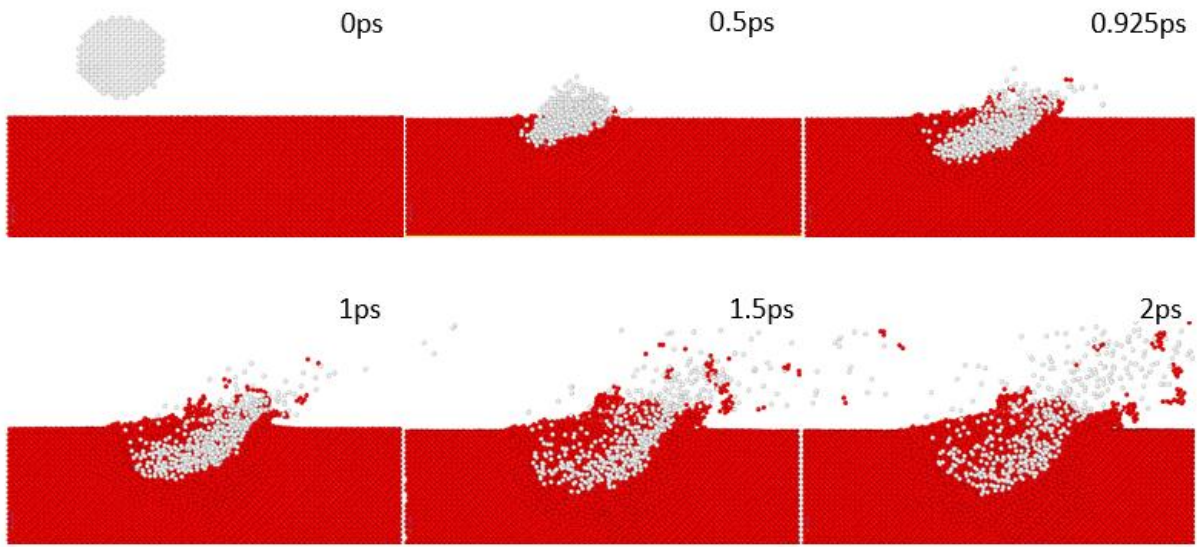

**Figure 3 Cross-section view of the time evolution of 20eV/atom $Ar_{1000}$ cluster ion impact on the Si(100) surface. Silicon atoms are shown in red, argon atoms in grey.**

These results are in general agreement with the sputtering mechanism for large Ar clusters at 45° angle of incidence suggested by Postawa et al[22]. The kinetic energy of the Ar clusters is deposited close to the surface which leads to initial sputtering, seen at 0.925ps , in Fig 3. The atoms 'slide' to the side of the crater and interact with substrate particles which leads to secondary sputtering phase where atoms are 'washed out' of the crystal, as shown at 1.5ps in Fig 3 . Fig. 4illustrates the evolution of the system at 2ps  for a range of energies. The crater size, depth of crater and sputtering yield all increase with the increase in the energy per atom of the Ar cluster, as might be expected.

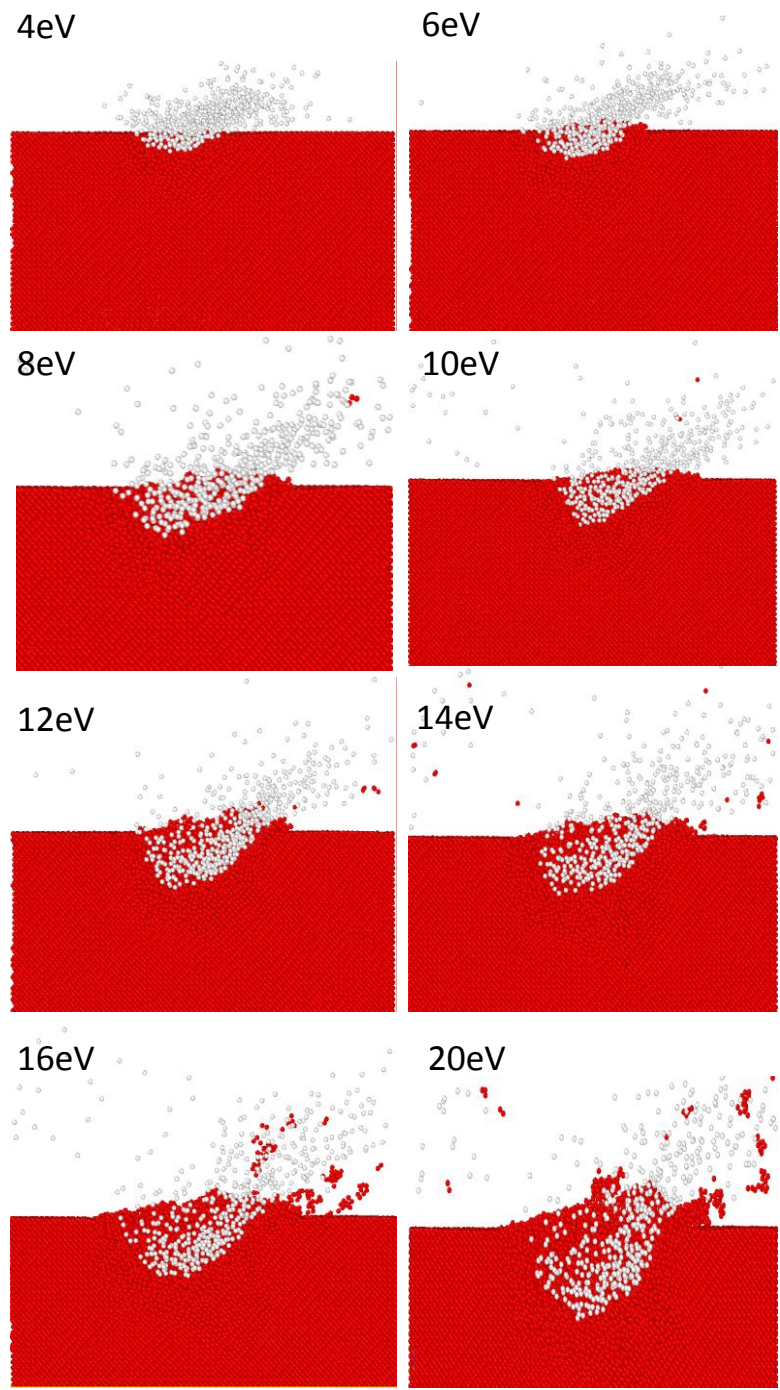

**Figure 4 Effect of the impact of a 1000 atom Ar cluster for energy-per-atom of 4, 6, 8, 10, 12, 14, 16 and 20 eV, each 2ps after impact. The cluster arrives at 45° to the surface from the left in these images. Argon atoms are in grey and silicon atoms in red.**

## 3.2 Quantitative description of MD results using the Threshold model

We performed 1,150 individual simulations of an argon cluster impacting the Si (100) surface. Each took around 2.5 hours to complete on a 40-core processing node of the "Topsy" HPC cluster at Newcastle University. The number of sputtered silicon atoms was counted automatically from the MD results, and the sputter yield estimated as the average number of sputtered atoms per cluster atom in the impacting cluster ion. The points shown in Fig. 5 represent these MD results.

In a previous work[27] we fitted experimental sputter yield data to the "Threshold" model, i.e. Eqn (1). This was almost all based on the use of nominally $n$=1,000 atom clusters. Since that work was published Paruch *et al*[16] have proposed, on the basis of MD calculations, a power law dependence of total sputter yield on the ratio of total cluster energy. Our own MD calculations show a similar trend to larger yields for larger clusters even when the energy-per-atom is the same. Therefore, to extend the range of Eqn (1) to describe the total sputter yield for a wide range of Ar cluster sizes we incorporate this power law exponent to Eqn (1) to give;

$$Y(\epsilon) = A' \, (n\epsilon)^\alpha \left[ 1 + \mathrm{erf}\left(\frac{\epsilon - U}{s}\right) \right]$$

(5)

Where $\varepsilon$ and $U$ are in eV. The continuous lines in Fig 5 represent the results of fitting Eqn(5) to our MD data, adjusting $A'$, $U$, $s$ and $\alpha$. We obtained U=13.2eV, $\alpha = 1.57$, s=4.9eV and A'=8.1x10$^{-7}$. All of these values are in the range we would expect. Wu et al[28] have reported a value between 18eV and 20eV for the sputter threshold energy from Si, but $U$ is an effective parameter and we would not expect exact agreement with measured values. What is more, experiments performed by Wu et al[28] were for normal incidence, the sputter threshold is likely to be slightly lower for angles of incidence commonly used in SIMS and XPS experiments. In previous experimental work we found that for inorganic materials $s \approx 0.5U$, whereas in this fit to MD results we have $s \approx 0.37U$. In experimental measurements it is likely that $s$ takes on a slightly higher value than in MD calculations because of the finite spread of cluster sizes from any real gas cluster source. In previous MD simulations Paruch *et al* [16] found $\alpha = 1.2$ for Au(111), Pt(111) and Cu(111) surfaces and the condition $1 < \alpha < 2$, at least for small clusters[29].

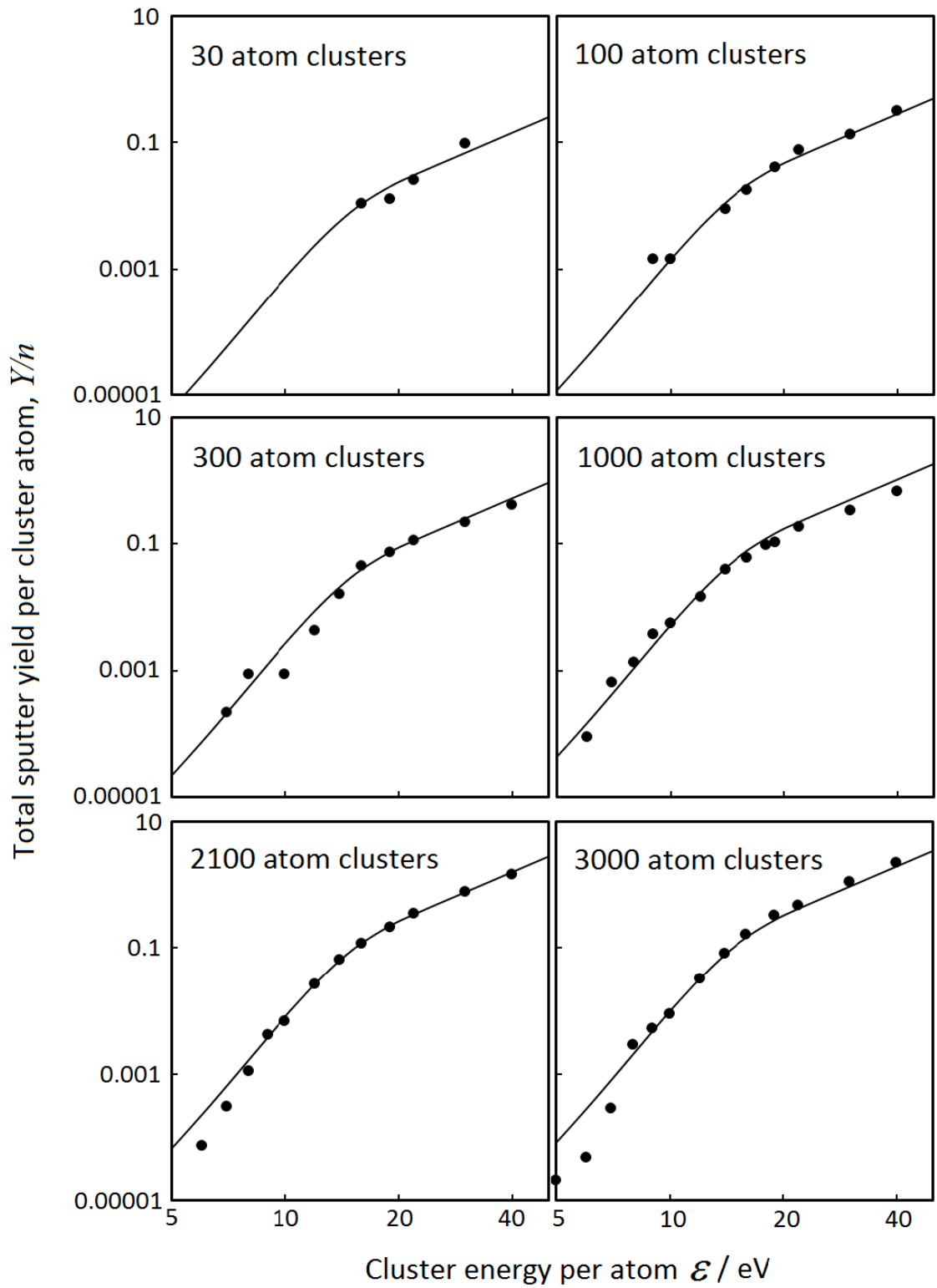

Figure 5 Total sputter yield from the Si(100) surface for six different Ar cluster ion sizes. The points are MD calculations, each point representing the average number of Si atoms sputtered per Ar atom in the cluster for between 7 and 56 separate MD simulations. The continuous lines represent the Threshold Equation model, Eqn(5), fitted to this data.

Eqn (5) therefore describes the MD results extremely well. For low energy/atom regime (below monatomic sputter threshold). The 'knee' of the graph occurs around the monotomic sputter threshold and indicates the change to a high energy-per-atom regime, where all of the atoms in the cluster have sufficient energy to cause sputtering.

## 4. Conclusions

We have demonstrated the effectiveness of LAMMPS, in combination with carefully-chosen potential functions, for the simulation of gas cluster ion impacts important in modern XPS and SIMS experimental depth-profiling.

A set of 1,150 Molecular Dynamics simulations of Ar cluster impacts on Si(100) have been conducted, ranging from 30 to 3000 Ar atoms, and 5eV to 40eV energy per atom. The existing "Threshold" model for total sputter yield estimation has been extended to this wider range of cluster size, and Eqn(5) confirmed as modelling this yield extremely well. Eqn (5) should prove extremely useful in future XPS and SIMS studies involving argon gas clusters.

**Acknowledgements**

The majority of this work was performed at the National EPSRC XPS User's Service (NEXUS) at Newcastle University, an EPSRC Mid-Range Facility. This was made possible, in part, by an instrument package funded by EPSRC's 'Great Eight' capital funding grant EP/K022679/1 and Newcastle University. MJ is grateful for receipt of a Rowe Textron Scholarship for 2016 during which this work was initiated in Cincinnati and continued in Newcastle.

# Appendix: Potentials and Constraints

## A1. Lennard-Jones potential

Lennard-Jones potential is a relatively simple mathematical function, which approximates the potential energy between two atoms. The function takes the form of a pair-wise interaction and is a function of distance between two particles.

$$V(r) = 4d\left[\left(\frac{\sigma}{r}\right)^{12} - \left(\frac{\sigma}{r}\right)^{6}\right]$$

(A1)

Where $d$ is the depth of the potential well, and σ is the distance at which the potential energy between the two atoms is zero. At short distances the term $\left(\frac{\sigma}{r}\right)^{12}$ dominates. It models the steep repulsion caused by overlapping of electronic clouds surrounding the atoms. At large distances the term $\left(\frac{\sigma}{r}\right)^{6}$ dominates and this accounts for van der Waals interactions. This weak interaction is a key bonding characteristic in noble gases, hence Lennard-Jones approximates the potential energy between Ar atoms reasonably well. The equation parameters have been fitted to reproduce experimental data, for Ar $d$=0.0104 eV and σ=3.40Å The values found by Michels *et al*[30], are close to the parameters used in our simulations. The cut off distance used in the simulation is, $r_c$=8Å.

## A2. Ziegler-Biersack-Littmark (ZBL) potential

Zigeler et al[31] produced a universal function which computes the screened nuclear repulsion energy for any atom pair. In our work it is used to model the interaction between Ar and Si atoms. The ZBL function is a two-body Coulomb screening potential which has become the standard for modelling high-energy collisions between atoms, it takes the form:

$$E_{ij}^{ZBL} = \frac{1}{4\pi\varepsilon_0}\frac{Z_i Z_j e^2}{r_{ij}}\Phi + S(r_{ij})$$

(A2)

Where $Z_i$ and $Z_j$ are the nuclear charges of the two atoms, $\varepsilon_0$ is the electrical permittivity of vacuum, e is the electron charge, $r_{ij}$ is the distance between the two atoms, $\Phi\left(\frac{r_{ij}}{a}\right)$ is the universal screening function and $S(r_{ij})$ is a switching function. The screening function is expressed in terms of $r_{ij}$ and scaling parameter a;

$$\Phi(x) = 0.18175e^{-3.19980x} + 0.50986e^{-0.94229x} + 0.28022e^{-0.40290x}$$
$$+ 0.02817e^{-0.20162x}$$
$$where\ a = \frac{0.46850}{Z_i^{0.23} + Z_j^{0.23}}, and\ x = \left(\frac{r_{ij}}{a}\right)$$

(A3)

The switching function is a third-order spline function, which acts between an inner and outer cut-off. It modifies the curvature of the energy plot and reduces the ZBL function value to zero at cut-off, $r_c$.

$$S(r) = C \quad\quad r < r_1$$
$$S(r) = \frac{A}{3}(r-r_1)^3 + \frac{B}{4}(r-r_1)^4 + C \quad\quad r_1 < r < r_c$$
$$A = \frac{(-3E'(r_c) + (r_c - r_1)E''(r_c))}{(r_c - r_1)^2}$$
$$B = \frac{(-2E'(r_c) - (r_c - r_1)E''(r_c))}{(r_c - r_1)^3}$$
$$C = -E(r_c) + \frac{1}{2}(r_c - r_1)E'(r_c) - \frac{1}{12}(r_c - r_1)^2 E''(r_c)$$

(A4)

Where E' and E'' are the first and second derivatives of Coulomb screening potential. Values specified in the simulations for the different variables are equal to: $Z_{Si} = 14, Z_{Ar} = 18$, $r_c = 8$Å and $r_1 = 5$Å.

A3. **Stillinger-Weber potential**

Three requirements must be fulfilled by a potential used to model the interaction between silicon atoms in our sputtering simulations. It must be able to describe the three phases: crystalline, amorphous and liquid equally well because they co-exist in collision cascades. The phase transition must occur at temperatures close to experimental values and the binding of atoms on the surface needs to be realistic. If the binding is too weak the characteristic crater rims will not form, and a higher sputtering yield will be observed than is true of real systems[32].

The Stillinger-Weber potential consists of both two-body and three-body terms. The three-body term strengthens the attraction between Si atoms in the direction of the favoured tetrahedral bond angle, of 109.5° [33]:

$$E = \sum_i \sum_{j>i} \Phi_2(r_{ij}) + \sum_i \sum_{j \neq i} \sum_{k>j} \Phi_3(r_{ij}, r_{ik}, \theta_{ijk})$$

$$\Phi_2(r_{ij}) = A_{ij} \epsilon_{ij} \left[ B_{ij} \left( \frac{\sigma_{ij}}{r_{ij}} \right)^{\wedge}(-p_{ij}) - \left( \frac{\sigma_{ij}}{r_{ij}} \right)^{\wedge}(-q_{ij}) \right] \text{EXP} \left( \frac{\sigma_{ij}}{r_{ij} - a_{ij}\sigma_{ij}} \right)$$

$$\Phi_3(r_{ij}, r_{ik}, \theta_{ijk}) = \lambda_{ijk} \epsilon_{ijk} \left[ \text{COS } \theta_{ijk} - \text{COS } \theta_{0ijk} \right]^2 \text{EXP} \left( \frac{\gamma_{ij}\sigma_{ij}}{r_{ij} - a_{ij}\sigma_{ij}} \right) \text{EXP} \left( \frac{\gamma_{ik}\sigma_{ik}}{r_{ik} - a_{ik}\sigma_{ik}} \right)$$

(A5)

Where, following Stillinger and Weber[33], A=7.049556277, B=0.6022245584, p=4,q=0, a=1.8, λ=21, γ=1.2, $\text{COS } \theta_{0ijk} = -\frac{1}{3}$, σ=2.0951 and ε= 2.1683 $eV$ The Stillinger-Weber potential models the three phases well and the melting temperature of crystalline structure closely matches experimental values[32].